\documentclass[final]{jpp}
\pdfoutput=1

\usepackage[T1]{fontenc}
\usepackage[utf8]{inputenc}
\usepackage[english]{babel} 
\usepackage{amsmath}
\usepackage{amssymb}
\usepackage{textcomp}
\usepackage{lmodern}

\usepackage{color}
\usepackage{graphicx}
\usepackage[hidelinks=true,
colorlinks=true,
linkcolor=blue,
citecolor=blue,
urlcolor=blue
]{hyperref}
\usepackage{natbib}

\graphicspath{ {figures/} }

\usepackage{physics}
\usepackage{units}
\usepackage[nosfdefault]{comicneue}
\newcommand{\Smilei}{{\comicneue Smilei}}

\newcommand{\micro}{\text{\textmu}}

\newcommand{\FWHM}{\text{FWHM}}
\newcommand{\ee}{\mathrm{e}}
\newcommand{\ii}{\mathrm{i}}

\newcommand{\nc}{n_{\mathrm{c}}}

\newcommand{\Cu}{{\mathrm{Cu}}}
\newcommand{\EE}{\mathcal{E}}

\newcommand{\jxB}{$j{\hspace{-0.4pt}\times\hspace{-0.5pt}}B$}

\usepackage{etoolbox}
\makeatletter

\patchcmd{\NAT@citex}
  {\@citea\NAT@hyper@{%
     \NAT@nmfmt{\NAT@nm}%
     \hyper@natlinkbreak{\NAT@aysep\NAT@spacechar}{\@citeb\@extra@b@citeb}%
     \NAT@date}}
  {\@citea\NAT@nmfmt{\NAT@nm}%
   \NAT@aysep\NAT@spacechar\NAT@hyper@{\NAT@date}}{}{}

\patchcmd{\NAT@citex}
  {\@citea\NAT@hyper@{%
     \NAT@nmfmt{\NAT@nm}%
     \hyper@natlinkbreak{\NAT@spacechar\NAT@@open\if*#1*\else#1\NAT@spacechar\fi}%
       {\@citeb\@extra@b@citeb}%
     \NAT@date}}
  {\@citea\NAT@nmfmt{\NAT@nm}%
   \NAT@spacechar\NAT@@open\if*#1*\else#1\NAT@spacechar\fi\NAT@hyper@{\NAT@date}}
  {}{}

\makeatother

\begin{document}%% v v v v v v v v v v v v v v v v v v v v v v v v v v
%%%%%%%%%%%%%%%%%%%%%%%%%%%%%%%%%%%%%%%%%%%%%%%%%%%%%%%%%%%%%%%%%%%%%%

% \doi{10.1017/S0022377820000264}
% \journal{J.\ Plasma Phys.}
% \pubyear{2020}
% \volume{86}
% \pagerange{755860201}

%%%%%%%%%%%%%%%%%%%% vvv Internal title page vvv %%%%%%%%%%%%%%%%%%%%%
\shorttitle{Fast electron heating and thermalization with CP laser}
\shortauthor{A.~Sundstr\"{o}m, L.~Gremillet, E.~Siminos and
  I.~Pusztai} %
\title{\vspace{2em}Fast collisional electron heating and relaxation
  in thin foils driven by a circularly polarized ultraintense
  short-pulse laser} %
\author{Andr\'{e}as Sundstr\"{o}m\aff{1}
  \corresp{\email{andsunds@chalmers.se}}, %
  Laurent Gremillet\aff{2}, %
  Evangelos Siminos\aff{3} %
  \and Istv\'{a}n Pusztai\aff{1}}
\affiliation{
  \aff{1}Department of Physics, Chalmers University of
  Technology, 41296 Gothenburg, Sweden
\aff{2} CEA, DAM, DIF, F-91297 Arpajon, France
  \aff{3}Department of Physics, University of Gothenburg,
  41296 Gothenburg, Sweden
}
\maketitle

\begin{abstract}
The creation of well-thermalized, hot and dense plasmas is attractive
for warm dense matter studies.
We investigate collisionally induced energy absorption of an
ultraintense and ultrashort laser pulse in a solid copper target using
particle-in-cell simulations. We find that, upon irradiation by a
$\unit[2\times10^{20}]{W\,cm^{-2}}$ intensity, $\unit[60]{fs}$
duration, circularly polarized laser pulse, the electrons in the
collisional simulation rapidly reach a well-thermalized distribution
with ${\sim}\unit[3.5]{keV}$ temperature, while in the
collisionless simulation the absorption is several orders of magnitude
weaker. Circular polarization inhibits the generation of suprathermal
electrons, while ensuring efficient bulk heating through inverse
bremsstrahlung, a mechanism usually overlooked at relativistic laser
intensity.
An additional simulation, taking account of both collisional and field
ionization, yields similar results: the bulk electrons are heated to
${\sim}\unit[2.5]{keV}$, but with a somewhat lower degree of
thermalization than in the pre-set, fixed-ionization case.
The collisional absorption mechanism is found to be robust against
variations in the laser parameters. At fixed laser pulse energy,
increasing the pulse duration rather than the intensity leads to a
higher electron temperature.
\end{abstract}

\begin{keywords}
plasma simulation, plasma heating, plasma dynamics
\end{keywords}

%%%%%%%%%%%%%%%%%%%% ^^^ Internal title page ^^^ %%%%%%%%%%%%%%%%%%%%%

\section{Introduction}

The creation of warm dense matter (WDM) or hot dense matter (HDM) in a
laboratory setting is of high interest for a broad field of research
disciplines such as laboratory
astrophysics~\citep{Remington_PPCF2005,Bailey-etal_PRL2007,Fujioka-etal_NPhys2009}, %
studies of planetary
interiors~\citep{Ross_Nat1981,Knudson-Etal_Sci2008}, %
inertial confinement
fusion~\citep{Drake_NF2018,LePape-etal_PRL2018}, %
understanding the equations of state under such extreme
conditions~\citep{Renaudin-etal_PRL2003,Nettelmann-etal_ApJ2008} %
and experimental verification of high energy density (HED) atomic
physics
models~\citep{Hoarty-etal_PRL2013,Faussurier-Blancard_PRE2019}. %
However, in order to benchmark atomic physics models against
spectroscopic data, these must be obtained under well-controlled
conditions. Since most such models assume Maxwellian plasma
populations, this means that, when diagnosed, the heated samples
should be as close to thermal equilibrium as possible.

The generation of WDM/HDM at uniform near-solid density requires that
the sample be heated rapidly, i.e.\ before any significant
hydrodynamic expansion. %
Such isochoric heating can be achieved using ultrahigh-intensity,
short-pulse lasers, as has been done at various high-power
systems~\citep{Evans-etal_APL2005,Gregori-etal_CTPP2005,Martinolli-etal_PRE2006,Chen-etal_PoP2007,Nilson-etal_PRE2009,Perez-etal_PRL2010,Brown-etal_PRL2011,Hoarty-etal_HEDP2013}.
These experiments were conducted using laser pulses with
$\unit[0.3{-}10]{ps}$ duration and energies in the range of
$\unit[10{-}500]{J}$, but there is a need for a wider access at
lower-energy table-top facilities, typically delivering joule-level,
femtosecond laser pulses. %
Promising results in this direction have recently been obtained by
\citet{Purvis-etal_NPhoton2013} and \citet{Bargsten-etal_SciAdv2017}
making use of nano-wire arrays to strongly enhance the laser-to-plasma
coupling efficiency, thus creating keV temperature, sub-solid density
plasmas. Yet, such structured targets are non-trivial to manufacture
and are extremely sensitive to parasitic laser prepulses, which can
destroy the nano-structures before the main pulse arrives.

Most laser-based isochoric heating experiments conducted so far have
exploited the fast electrons driven by a linearly polarized laser
pulse~\citep{Nilson-etal_PRL2010,Santos-etal_NJP2017,Sawada-etal_PRL2019}. %
Their energy dissipation through the plasma bulk enables heating to
high temperatures ($\unit[0.1{-}1]{keV}$) at solid-range plasma
densities, but usually at the expense of poor spatial
uniformity~\citep{Dervieux-etal_HEDP2015} and relatively slow
thermalization. %
Plasma heating in this case is caused by the interaction of the fast
electrons with the bulk plasma via a combination of direct
collisions~\citep{Robinson-etal_NF2014}, ohmic dissipation of the
colder return
current~\citep{Lovelace-Sudan_PRL1971,Guillory-Benford_PP1972,Bell-Kingham_PRL2003,Robinson-etal_NF2014}
or plasma waves driven by the fast
electrons~\citep{Sherlock-etal_PRL2014}.
Some experiments have been done with laser-accelerated ions to heat a
secondary
target~\citep{Patel-etal_PRL2003,Dyer-etal_PRL2008,Mancic-etal_PRL2010}. Yet,
while this heating method can provide better spatial uniformity, it
leads to much lower ($\unit[{\sim}10]{eV}$) temperatures.

At normal laser incidence and linear polarization (LP), and for
sharp-gradient, highly overdense plasmas, the most commonly invoked
mechanisms of laser energy conversion into fast electrons are \jxB{}
heating~\citep{Kruer-Estrabrook_PoF1985} and vacuum
heating~\citep{Bauer-Mulser_PoP2007,May-etal_PRE2011}.
Both mechanisms hinge on the temporal modulation of the laser
ponderomotive force around the target surface, and thus lead to 
periodic injection of MeV range electron bunches into the target at
twice the laser frequency.
Such suprathermal electrons thermalize relatively slowly
(${\sim}\unit{ps}$), which may hinder those applications that require
a closely Maxwellian dense plasma.
In a work by \citet{Kemp-Divol_PoP2016}, it is shown that the fast
electron bunches induce surface waves that can scatter the energized
bulk electrons, thereby improving absorption. They also show the
necessity of collisions to first heat up the target surface to
$\unit{keV}$ temperatures, required for vacuum heating to commence. %

By contrast, in laser pulses with circular polarization (CP), for which
the ponderomotive force does not show high-frequency oscillations, the
\jxB{} and vacuum heating mechanisms are essentially suppressed in
overdense targets, and so is the fast electron bunch production (and
the surface waves induced by them). Still, some fast electrons can be
produced with CP if the variation time scale of the laser envelope is
not large compared to the laser
cycle~\citep{Siminos-etal_PRE2012,Siminos-etal_NJP2017}.

In this paper, we study the effects of collisions on the energy
absorption capability of the electrons in a thin, solid foil of a
high-atomic-number element. Due to the high atomic number, it is not
clear \emph{a priori} what degree of ionization ($Z^{*}$) the ions
have throughout the process and what influence the ionization history
has on heating. While a high $Z^{*}$ is desirable for the collisional
heating process, the initially cold target will not be highly ionized
in the beginning. Therefore, we have studied both different degrees of
fixed ionization as well as the self-consistent ionization process
including both field and impact ionization.

We demonstrate that the energy absorption of an intense short laser
pulse in a high-$Z^*$ solid-density target is mainly due to inverse
bremsstrahlung electron heating within the plasma skin layer, and that
this scenario holds in a broad range of experimentally relevant
parameters.
The front-layer electrons are collisionally scattered into the target
body where they heat the plasma bulk to $\unit{keV}$ level
temperatures, enough to reach $\unit{Gbar}$ range pressures, which is
well in the regime of HDM. %
The scattered electrons have sufficiently low energies that they
primarily heat the bulk via direct collisional thermalization. %
Since this mechanism relies on the scattering of the electrons
accelerated by the laser field against the heavy ions, it is operative
regardless the polarization. Inside the plasma, where the laser field
is negligible, collisions cause fast relaxation of the electron
distribution to a Maxwellian.

\section{Simulation design}
\label{sec:simulation}
We have performed one- and two-dimensional (1-D and 2-D respectively)
particle-in-cell (PIC) simulations of laser--solid interactions with
and without collisions enabled. We have used the \Smilei{} PIC
code~\citep{Smilei-paper}, which has a relativistic binary collision
module~\citep{Perez-etal_PoP2012} based on the collisional algorithm
by \citet{Nanbu_PRE1997} and \citet{Nanbu-Yonemura_JCompPhys1998}.  In
the case of a collisional plasma, we have considered either a fixed
degree of ionization or self-consistent modelling of the ionization
process~--~through field ionization and collisional impact ionization.

We ran 1-D simulations in a box of size $\unit[20]{\micro{m}}$ with a
resolution of $\rmDelta{x}=\unit[0.39]{nm}$ (51\,200
cells). We considered both LP and CP laser pulses with wavelength
$\lambda=\unit[800]{nm}$, dimensionless amplitude\footnote{%
  Note that the amplitude is normalized such that the intensity stays
  the same regardless of ellipticity, i.e. the field amplitude, at the
  same $a_0$ with circular polarization is
  $E_{\rm CP}=E_{\rm LP}/\sqrt{2}$ compared to that of LP,
  $E_{\rm LP}$. } %
$a_0=10$ (intensity
$I=\tfrac{1}{2}c\epsilon_0(m_{\ee}c\omega a_0/e)^2
\approx\unit[2\times10^{20}]{W\,cm^{-2}}$, where $\epsilon_0$ is the
vacuum permittivity, $m_{\ee}$ the electron mass, $e$ elementary
charge and $\omega$ the laser frequency) %
and a Gaussian temporal profile with $t_{\FWHM}=\unit[60]{fs}$
full-width-at-half-maximum (FWHM) duration in the intensity. The
plasma is $2.5{\rm\,\micro{m}}$ thick, starting at
$x=\unit[7.5]{\micro{m}}$ with a linear density ramp over a distance
of $\unit[20]{nm}$. The plasma consists of electrons and copper ions
at solid density,
$n_{\Cu,0}=48.4\nc\approx\unit[8.4\times10^{22}]{cm^{-3}}$, with
400~macro-particles per cell for each species. Here,
$\nc=\epsilon_{0}m_{\ee}\omega^{2}/e^2$ is the critical density
associated with the laser frequency $\omega$.  The particles are
initialized from Maxwell-J\"{u}ttner distributions (in three momentum
dimensions) with temperatures $T_{\ee,0}=\unit[1]{eV}$ for the
electrons and $T_{\ii,0}=\unit[0.1]{eV}$ for the ions. %

In order to assess the influence of the plasma collisionality alone,
we have first carried out simulations with fixed ionization degrees
$Z^*=$~11, 19, 24 and 27.  Then, to ascertain the physical accuracy of
these results, we have performed simulations describing both
collisional and field ionization.  The collisionless skin depth
$l_{\rm s}=c/[\omega(n_{\ee}/\nc)^{1/2}]$ is resolved, even for the
highest ionization where
$l_{\rm s}^{(Z^{*}=27)}\approx\unit[3.5]{nm}$.  The values of
$Z^{*}=11$, 19 and 27 correspond to full depletion of different
electronic shells, $Z^{*}=27$ being the reference ionization used in
other scans. An additional data point, $Z^{*}=24$ was chosen as an
arbitrary value between 19 and 27. When modelling the ionization
process self-consistently, the ions were initialized with $Z^{*}_0=5$,
in accordance with the widely used Thomas--Fermi
model~\citep{More_book1983}. Both field-tunnelling and electron--ion
impact ionization were enabled. The self-consistent ionization
simulation was only performed with CP.

We also performed one collisional 2-D simulation to check that our
results are robust to multidimensional effects.  This simulation uses
the same CP laser and target parameters as our 1-D base case. In order
to limit the computational cost at the increased dimensionality, it
was performed at a reduced resolution of 640 cells per micron in both
directions
($\rmDelta{x}=\rmDelta{y}=\unit[1.56]{nm}$), and a
simulation box size of $\unit[10]{\micro{m}}$ longitudinally and
$\unit[1.6]{\micro{m}}$ ($2\lambda$) transversely. Furthermore, the
number of particles per cell per species was reduced to $50$. A test
of these resolution parameters in one dimension showed excellent
agreement in electron kinetic energy spectrum of the main body of the
electrons with the corresponding high-resolution, collisional
\mbox{1-D} simulation; however, the lower particle count led to a
poorer statistics in the high-energy tail of the electron spectrum.

\section{Results and discussion}

\begin{figure}
\centering
\includegraphics{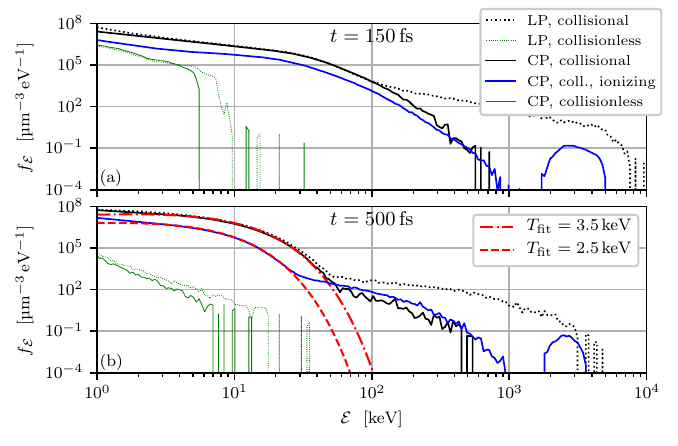}
\caption{Electron energy spectra $f_{\EE}$ at times
  $t=\unit[150]{fs}$~(a) and $t=\unit[500]{fs}$~(b), for LP (dotted
  lines) and CP (solid lines), with (black lines) and without
  collisions (thin, green lines); also showing CP with self-consistent
  field and collisional ionization (blue, solid line). %
  Two Maxwellian-J\"{u}ttner distributions are fitted to the bulk of
  the CP self-consistent and fixed ionization electron spectra in
  panel~b (red dashed and dash-dotted lines respectively).  }
\label{fig:energy-spectrum}
\end{figure}

Figure~\ref{fig:energy-spectrum} compares the electron energy spectra
as obtained at CP and LP ($Z^{*}=27$), with or without collisions
enabled and at CP with self-consistent ionization. The spectra are
recorded at two successive times ($t=\unit[150]{fs}$ and
$t=\unit[500]{fs}$)~--~for reference, the peak laser intensity hits
the target at $t\approx\unit[110]{fs}$ and the pulse FWHM duration is
$\unit[60]{fs}$. For both types of polarization, much higher electron
energies are achieved when allowing for collisions. %
The collisionally enhanced absorption results in a bulk electron
temperature of $T_{\ee}\approx\unit[3.5]{keV}$ at $t=\unit[500]{fs}$
for both LP and CP, determined by fitting Maxwell--J\"{u}ttner
distributions to the bulk spectra (i.e., ignoring the tails). %
Meanwhile, the collisionless simulations only reach an electron
temperature of ${\sim}\unit[10{-}100]{eV}$; these electrons are,
however, far from being thermalized and only their energetic tails are
visible in the figure.
The fact that both CP and LP reach very similar bulk electron
temperatures when collisions are enabled indicates that the laser
absorption mechanism is the same in both cases.

By comparing the electron spectra with the fitted Maxwellians, we
conclude that the electrons have reached a degree of thermalization
wherein less than $0.1\%$ of the kinetic energy is in the high-energy
tail for CP, and ${\sim}1\%$ for LP. The time frame for
this thermalization is consistent with a rough estimate for
electron--electron collisional energy transfer rates. At
$t=\unit[500]{fs}$, the deviation from Maxwellian starts at an
electron energy $\EE\simeq\unit[50]{keV}$. This is consistent with the
${\sim}\unit[300]{fs}$ collisional energy relaxation time of a
$\unit[50]{keV}$ electron through a $\unit[3.5]{keV}$ plasma at
$\unit[2.3\times10^{24}]{cm^{-3}}$ electron density~\citep{NRL2016}.

Note that these results compare LP and CP for the same on-target
intensity, while in an experiment, the circularly polarized pulse
would in practice be at a somewhat lower intensity due to losses in
the conversion from LP to CP (through a quarter-wave plate). Another
practical concern may be elliptical polarization, due to imperfections
in the CP conversion. While the exact dependence of the fast electron
generation on the ellipticity of the polarization is non-trivial, the
bulk collisional absorption itself is not expected to be affected by
the ellipticity, as demonstrated by the same bulk temperatures reached
in the two extreme cases of CP and LP.

The electron temperatures we quote are technically calculated before
the electrons have fully thermalized with the ions, which occurs over
${\sim}\unit{ps}$ time scales.  However, due to the high degree of
ionization, the electron--ion equilibrium temperature is
$T_{\rm eq}\approx{}n_{\ee}/(n_{\ee}+n_{\ii})T_{\ee}
\approx0.96T_{\ee}$. %
Thus, energy transfer from the electrons to the ions is insignificant.

\begin{figure}
\centering
\includegraphics{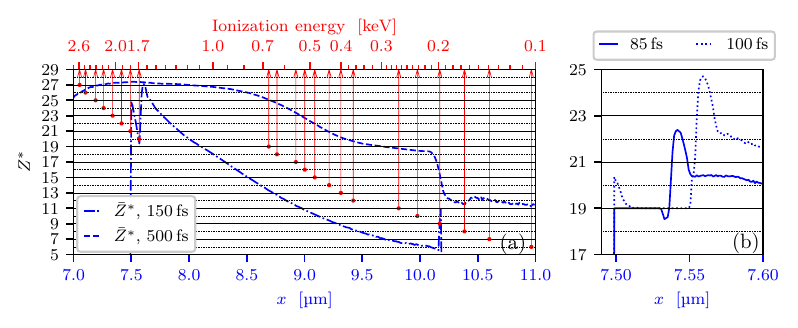}
\caption{Average ionization level profiles $\bar{Z}^{*}$ in the
  self-consistent field and collisional ionization simulation (blue
  lines, bottom axis) at times $t=\unit[150]{fs}$ (dash-dotted line)
  and $t=\unit[500]{fs}$ (dashed line), and ionization level as a
  function of ionization energies of copper~(red dots and arrows, top
  axis)~--~ionization data obtained from the Atomic Spectra Database
  of the National Institute of Standards and
  Technology~\citep{NIST_AtomicSpectraDatabase}. Panel~b shows the
  average ionization level profiles $\bar{Z}^{*}$ near the target
  front at $t=\unit[85]{fs}$ (solid line) and $t=\unit[100]{fs}$
  (dotted line).  }
\label{fig:avg_Zstar}
\end{figure}

When activating self-consistent (both field and collisional)
ionization, the bulk electron temperature is slightly reduced
(${\sim}\unit[2.5]{keV}$) compared to the fixed-ionization case. The
front plasma is rapidly ionized, mostly through field ionization, so
that collisional absorption quickly reaches an efficiency similar to
that obtained with fixed $Z^*=27$ (see
figure~\ref{fig:avg_Zstar}(\textit{b}) showing that the average
ionization $\ev{Z^{*}}\simeq24$ at the plasma front already at
$t=\unit[100]{fs}$). %
The lower $T_{\ee}$ is mostly due to the energy spent on
ionization~--~the average ionization energy from $Z^{*}=5$ to $27$ is
$\unit[0.9]{keV}$.

Moreover, figure~\ref{fig:energy-spectrum}(\textit{b}) shows that, for
both CP and LP, collisions cause efficient bulk electron
thermalization as early as $t=\unit[500]{fs}$. High-energy tails are
found to emerge above ${\sim}\unit[50]{keV}$ for the fixed ionization
and ${\sim}\unit[30]{keV}$ for the self-consistent ionization. Note
the large range of the logarithmic $f_{\EE}$ scale, meaning that the
tails are three to five orders of magnitude lower than the bulk
spectra.  The non-thermal tail is heavier in LP than in CP, due to the
operative \jxB{} and vacuum heating.

Also, the simulation with self-consistent ionization displays a larger
tail, compared to the bulk spectrum, than its counterpart with fixed
$Z^{*}$. %
The larger tail as well as an electron population at
${\sim}\unit[3]{MeV}$ can be explained by field-ionization events in
the charge-separation layer, which is exposed to stronger laser
fields. %
As the target front electrons are being pushed back by the
ponderomotive force, the ions remaining in the charge-separation layer
experience the less shielded laser field which quickly ionizes them
further. Since these newly freed electrons are injected into regions
of stronger laser fields, they are energized similarly to vacuum
heating in LP, thus resulting in a larger population of non-thermal
electrons, which, as in LP, thermalize relatively slowly.
Furthermore, the average ionization level is lower inside the target
with self-consistent ionization, as seen in
figure~\ref{fig:avg_Zstar}, thus reducing the efficacy of collisional
thermalization. Both these effects act to give a larger high-energy
tail.

The ionization level profiles ($\bar{Z}^{*}$) of the self-consistent
ionization simulation shown in figure~\ref{fig:avg_Zstar}(\textit{a})
represent the local ionization levels averaged over all
macro-particles in each spatial cell. Comparing the $t=\unit[150]{fs}$
(dashed-dotted line) and the $t=\unit[500]{fs}$ (dashed line) average
ionization curves, we see that the target front quickly reaches a high
ionization degree while the bulk is ionized more gradually. Since
there are no strong electric fields inside the plasma, the ionization
of the bulk must be driven by impact ionization.  The $\bar{Z}^{*}$
curve at $t=\unit[500]{fs}$ (dashed line) in
figure~\ref{fig:avg_Zstar}(\textit{a}) displays plateaus at
$\bar{Z}^{*}=27$, 19 and to some extent 11.  These plateaus result
from the large jumps in ionization energies between the successive
outermost electron shells (e.g., between $\bar{Z}^*=11$ and $12$, or
$\bar{Z}^*=19$ and $20$; see figure~\ref{fig:avg_Zstar}\textit{a}).
The ionization energy to reach $\bar{Z}^* = 28$ is
${\sim}\unit[11]{keV}$

Figure~\ref{fig:avg_Zstar}(\textit{b}) shows $\bar{Z}^{*}$ at the
target front surface ($x=\unit[7.5]{\micro{m}}$). At time
$t=\unit[85]{fs}$ (solid line), the ionization level has saturated at
$\bar{Z}^{*}=19$ due to the jump in ionization energy after
$Z^{*}=19$. Later, at $t=\unit[100]{fs}$ (dotted line), the laser
field has become strong enough to sustain field ionization beyond
$Z^{*}=19$, yielding the peak in $\bar{Z}^{*}$ near
$x=\unit[7.5]{\micro{m}}$.  Apart from the laser field, the
electrostatic field ($E_x$) induced by the laser ponderomotive force
at the target front causes additional ionization. This results in the
$\bar{Z}^{*}$ peak seen around $x=\unit[7.55]{\micro{m}}$, which moves
into the plasma as the charge-separation layer is pushed forward by
the laser ponderomotive force.

\begin{figure}
\centering
\includegraphics{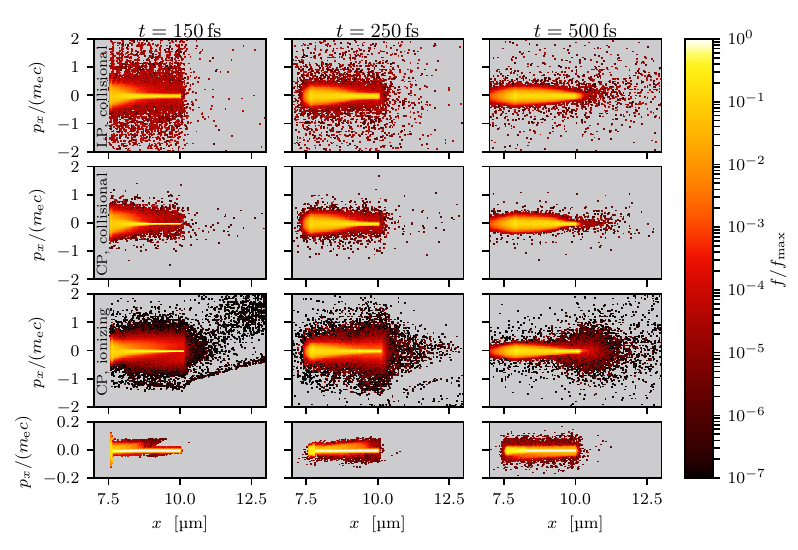}
\caption{Electron phase-space distributions for collisional LP with
  fixed ionization (top row); collisional CP with fixed ionization
  (second row) and self-consistent field and collisional ionization
  (third row) as well as for collisionless CP (bottom row), at times
  $t=\unit[150]{fs}$~(left column), $\unit[250]{fs}$~(middle column)
  and $\unit[500]{fs}$~(right column). %
  Note the different momentum scale for the collisionless CP.
}
\label{fig:electron_dist}
\end{figure}

The difference between the simulations is made clearer when studying
the electron phase spaces shown in figure~\ref{fig:electron_dist}%
\footnote{The normalization $f/f_{\max}$ of the distribution functions
  in figure~\ref{fig:electron_dist} and \ref{fig:py-pz} are with
  respect to the \emph{initial} maximum value of the distribution
  function in their respective planes of phase space $f_{\max}$. The
  colour values of the plotted distributions can therefore be directly
  compared.}.
The figure displays time sequences of the collisional distributions
with LP in the top row and CP in the second row; the third row shows
the self-consistent ionization CP simulation and the bottom row shows
the collisionless CP distribution. In the LP simulation, high-energy
electron bunches are produced at twice the laser frequency, as seen in
the $t=\unit[150]{fs}$ panel (top row), while CP with fixed ionization
(second row) produces a more even distribution of hot electrons since
\jxB{} and vacuum heating mechanisms are inhibited. At
$t=\unit[500]{fs}$, most of the fast electrons have thermalized in the
fixed-ionization case, while there remains a significant population of
high-energy electrons ``swarming'' around the back of the target with
self-consistent ionization.

With self-consistent ionization, two populations of relatively
high-energy electrons are created during the rising phase of the laser
pulse.
These populations originate from two successive field-ionization
phases. The first one occurs early in the interaction, when the
ionization of the surface plasma momentarily saturates at
$\bar{Z}^{*}=19$. In the $t=\unit[150]{fs}$ panel of
figure~\ref{fig:electron_dist}, this population accounts for the broad
momentum distribution in the target bulk, and also for the beam (with
momenta $p_x/(m_{\ee}c) \sim -1$) being reflected in the vacuum
($x>\unit[10]{\micro{m}}$) and refluxing into the target.  The second
phase starts at $t\simeq\unit[90]{fs}$, when the laser pulse gets
intense enough to ionize the surface plasma beyond $Z^{*}=19$ (compare
the $\unit[85]{fs}$ and $\unit[100]{fs}$ curves in
figure~\ref{fig:avg_Zstar}\textit{b}). This yields fast electrons
(visible in the upper right corner of the $t=\unit[150]{fs}$ phase
space) more energetic than those generated earlier, which correspond
to the bump around $\sim 3\,\rm MeV$ in the energy spectra of
figure~\ref{fig:energy-spectrum}.
A similar field-ionization injection of fast electrons from the
surface ions was seen by \citet{Kawahito-Kishimoto_PoP2017} in a
carbon plasma, although they used LP which also caused bunching of the
electrons at twice the laser frequency.

\begin{figure}
\centering
\includegraphics{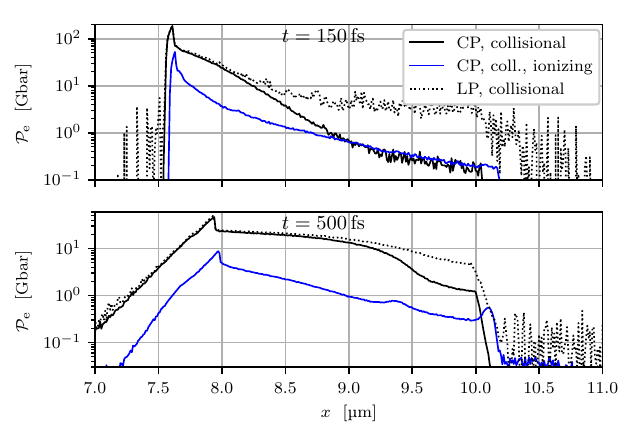}
\caption{Electron kinetic energy density $\mathcal{P}_{\ee}$ for
  collisional LP (dotted line) and CP (black solid line) and CP with
  self-consistent field and collisional ionization (blue solid
  line), at times $t=\unit[150]{fs}$ (top panel) and
  $t=\unit[500]{fs}$ (bottom panel). The peak laser intensity hits the
  target at $t\approx\unit[110]{fs}$. }
\label{fig:electron_E-dens}
\end{figure}

We now turn our attention to the energy density achieved in these
scenarios. Since the heating process is fast compared to
hydrodynamical time scales, the plasma bulk has not had time to
expand, and hence the bulk electrons and ions remain at solid-range
density. At the same time, the electrons reach keV temperatures,
resulting in high energy densities of the order of
${\sim}\unit[10]{Gbar}=\unit[10^{9}]{J/cm^{3}}$. In
figure~\ref{fig:electron_E-dens}, the electron kinetic energy density
$\mathcal{P}_{\ee}$ is displayed throughout the target at times
$t=\unit[150]{fs}$ (top panel) and $t=\unit[500]{fs}$ (bottom
panel). At the earlier time, the kinetic energy density is
concentrated to the front of the target, while at the later stage the
energy has spread out throughout most of the plasma. The energy
density in the fixed-ionization simulations reaches approximately
${\sim}\unit[10]{Gbar}$ and is mostly homogeneous in the region
$x=\unit[8{-}9]{\micro{m}}$.

The high-energy electrons created with LP facilitate a better spatial
homogenization of the energy density than with CP. Their slow
thermalization results in a more spatially homogeneous target heating,
since they can recirculate several times through the plasma. In a
potential application, one should therefore make a compromise between
good thermalization and homogenization. Another parameter that can be
used to control homogenization is target thickness; decreasing it
helps for a faster homogenization of the plasma heating. However, a
thinner target will also explode faster hydrodynamically, which would
give a HED application a shorter time frame to operate in.

Meanwhile, the self-consistent simulation gives an exponentially
decreasing energy density profile throughout the target at
$t=\unit[500]{fs}$, indicating that thermalization is taking
longer. The lower temperature and electron density reached with
self-consistent ionization result in approximately an order of
magnitude lower energy density compared to the fixed-ionization
results. %
However, there is still a significant region with
$\mathcal{P}_{\ee}>\unit[1]{Gbar}$ in the self-consistent ionization
simulation at $t=\unit[500]{fs}$. In this case, the energy density
does not homogenize as efficiently, partly due to a decreased ability
of the target to thermalize fast electrons (stemming from lower
$\bar{Z}^{*}$), and partly due to the inhomogeneity of the ionization
profile which affects the bulk electron density profile.

As a consequence of the strong gradients in $\mathcal{P}_{\ee}$ around
the target front side, a shock wave is launched. The shock wave
presents itself as a sharp jump in electron pressure, most clearly
seen close to $x=\unit[8.0]{\micro{m}}$ in the $t=\unit[500]{fs}$
panel in figure~\ref{fig:electron_E-dens}.  The details of shock
formation are sensitive to the laser and target parameters, and are
more clearly seen from the ion phase space, as will be addressed by a
paper in preparation~\citep{IonPaper2020}.  However, no ion reflection
occurs at the shock front, which means that the shock is hydrodynamic
like in its nature.

At the high ionization levels discussed in this paper, such high
temperatures and densities may result in significant energy losses due
to bremsstrahlung. The total bremsstrahlung emission power density can
be estimated as
$S_{\rm BS}\,[\unit{W\,cm^{-3}}]\approx 1.69{\times}10^{32}\times
{Z^{*}}^{3} (n_\ii\,[\unit{cm^{-3}}])^{2}
\,(T_{\ee}\,[\unit{eV}])^{1/2}$~\citep{NRL2016}. %
By comparing the power density to the thermal energy density
${\sim}\frac{3}{2}n_{\ee}T_{\ee}$, we arrive at a radiative time scale
of the order of several $\unit{ps}$ for keV range temperatures at a
density of $n_{\ii}=\unit[8.4\times10^{22}]{cm^{-3}}$. Hence, the
radiative losses from bremsstrahlung will mostly be of concern at time
scales longer than those studied in this paper. However,
bremsstrahlung losses cannot be completely disregarded in a WDM/HDM
experiments, where current spectroscopic temporal resolution is
constrained to ${\sim}\unit{ps}$ time scales.

Besides bremsstrahlung, line emission from relaxation of excited
states may be of concern. However, at electron energies above
$\gtrsim\unit[10]{eV}$, electron energy losses from excitations become
subdominant compared to
ionization~\citep{Joshipura-etal_Pramana2006}. Since energy loss from
collisional ionization events is accounted for in the self-consistent
ionization simulation, the temperature of
$T_{\ee}\approx\unit[2.5]{keV}$ is likely not affected much by losses
through line emissions.

\subsection{Illustration of the
  collisional absorption mechanism} %

To illustrate the mechanism by which collisions enhance absorption, we
have performed a simplified set of simulations. These are designed to
generate a quasi-steady state: the laser intensity is constant after a
linear ramp-up over 10~laser cycles; the ions are stationary; the
plasma is $\unit[2.5]{\micro{m}}$ long and it terminates at a thermal
boundary, meaning that particles which exit the boundary are reflected
with momenta chosen randomly from a Maxwellian distribution at
$T_{\ee,0}=\unit[10]{eV}$ for the electrons~--~the same at the initial
temperatures. The other simulation parameters are: CP at $a_0=10$,
$Z^{*}=27$ with and without collisions; resolution and other numerical
parameters are as stated in \S~\ref{sec:simulation}. The long ramp-up
time has been chosen to reduce electron energization due to the laser
amplitude envelope modulation~\citep{Siminos-etal_PRE2012}. %
Note that due to the steady state nature of this simplified simulation
setting, it is hard to draw any quantitative conclusions that can be
transferred to the time-varying situation.

We will now take a look at the interaction between the electrons and
the laser electric field. The density of power $S$ exerted on an
electron population can be expressed as
\begin{equation}\label{eq:S}
S(x,t)
=-e\int\!\dd[3]v\,\vb*{E}_{\perp}\vdot\vb*{v}f_{\ee}(\vb*v)
={-}en_{\ee}\vb*{E}_{\perp}\vdot\vb*{V}_{\perp},
\end{equation}
where $\vb*{E}_{\perp}=\vb*{E}_{\perp}(x,t)$ is the laser electric
field~--~which only lies in the transverse plane~--~and
$\vb*{V}_{\perp}=\vb*{V}_{\perp}(x,t)\equiv
[1/n_{\ee}(x,t)]\int\!\dd[3]v\,\vb*{v}_{\perp}f_{\ee}(x,\vb*v;t)$ %
is the projection of electron velocity moment onto the transverse
plane.

In a 1-D model, disregarding collisional effects, the transverse
canonical momentum
$\tilde{\vb*{P}}_\perp=\vb*{P}_{\perp}-e\vb*{A}_{\perp}$ is conserved,
and $\tilde{\vb*{P}}_\perp=0$. Hence
$\vb*{P}_{\perp}=e\vb*{A}_{\perp}$, where $\vb*{A}_{\perp}$ and
$\vb*{P}_{\perp}$ are the transverse component of the magnetic vector
potential and the electron momentum moment, defined analogously to
$\vb*{V}_{\perp}$. %
In quasi-steady state, $\vb*{A}_{\perp}$ is just rotating in the
transverse plane, so the electric field is
$\vb*{E}_{\perp}\equiv-\pdv*{\vb*{A}_{\perp}}{t} =\omega A_{\perp}
[\cos(\omega{t})\vu*{y}-\sin(\omega{t})\vu*{z}]/\sqrt{2}$, %
where $A_{\perp}=A_{\perp}(x)$ is the magnitude of the vector
potential (necessarily transverse in 1-D). Importantly, the electric
field vector is perpendicular to the vector potential and the
magnitude of the electric field is $E_{\perp}=\omega{}A_{\perp}$. We
therefore expect $\vb*{P}_{\perp}$ and $\vb*{E}_{\perp}$ to be
perpendicular and their magnitudes~--~in normalized units~--~to be
equal, ${P}_{\perp}={E}_{\perp}$.

\begin{figure}
\centering
\includegraphics{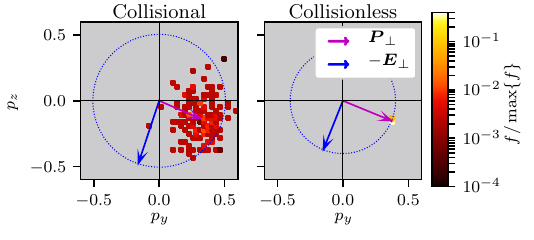}
\caption{Electron transverse momentum distribution at
  $x=\unit[7.0{-}7.2]{nm}$ behind the ion front in the simplified
  simulations with fixed ions. The left and right panels correspond to
  collisional and collisionless simulations, respectively.  The
  distributions are here recorded at $t=\unit[190]{fs}$, which is well
  after the quasi--steady state has been reached, where
  $\vb*{E}_{\perp}$ and $\vb*{P}_{\perp}$ rotates (clock-wise) in the
  transverse plane. }
\label{fig:py-pz}
\end{figure}

\begin{figure}
\centering
\includegraphics{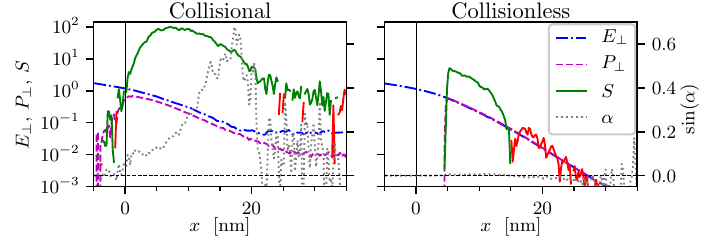}
\caption{Magnitudes of the perpendicular electric field $E_{\perp}$
  (blue dash-dotted) and mean electron transverse momentum $P_{\perp}$
  (magenta dashed) as well as the absorbed power density $S$ (solid
  line, green and red for $S>0$ and $S<0$ respectively). Also shown is
  the phase shift $\sin(\alpha)$ (grey dotted) between
  ${-}\vb*{E}_{\perp}$ and $\vb*{V}_{\perp}$, where $\vb*{V}_{\perp}$ is
  the mean electron transverse velocity moment of the
  distribution. The vertical black line marks the location of the
  transverse momentum planes plotted in figure~\ref{fig:py-pz}. %
  All values are expressed in dimensionless units.  }
\label{fig:perp}
\end{figure}

Figure~\ref{fig:py-pz} shows slices of the collisional (left) and
collisionless (right) electron distributions in the transverse
momentum plane at $t=\unit[190]{fs}$, well after quasi-steady state
has been reached, and in a thin slice $\unit[7.0{-}7.2]{nm}$ (one cell
length) behind the immobile ion front edge of the plasma.
If we were to evolve this picture in time, we would see the (negative)
electric field ${-}\vb*{E}_{\perp}$ rotate clockwise, along the
marked-out circular path in figure~\ref{fig:py-pz}; the mean momentum
$\vb*{P}_{\perp}$ would follow synchronously in this rotation.
The most apparent difference between the collisional and collisionless
distributions is the much larger momentum spread of the former, caused
by collisional scattering of the electrons.
In contrast to the collisionless case, ${-}\vb*{E}_{\perp}$ and
$\vb*{P}_{\perp}$ are not equal in magnitude nor are they perfectly
perpendicular. The missing transverse canonical momentum has been
collisionally transferred to the ions, where it disappears from the
simulation due to the ions being static.
Note that if ${-}\vb*{E}_{\perp}$ and $\vb*{P}_{\perp}$ are not
perfectly perpendicular in \eqref{eq:S}, then the absorbed power
density $S$ is non-vanishing\footnotemark{}. %
\footnotetext{For simplicity, we are ignoring relativistic effects in
  this discussion, which would otherwise complicate the relationship
  between $\vb*{P}_{\perp}$ and $\vb*{V}_{\perp}$.} %
We can express \eqref{eq:S} as
\begin{equation}
S={-}ne\vb*{E}_{\perp}\vdot\vb*{V}_{\perp}
=ne E_{\perp} V_{\perp}\sin(\alpha),
\end{equation}
where the phase angle between ${-}\vb*{E}_{\perp}$ and $\vb*{V}_{\perp}$ is
$\upi/2-\alpha$. %

Figure~\ref{fig:perp} displays configuration space profiles of
$E_{\perp}$, $V_{\perp}$ and $S$~--~in dimensionless units~--~as well
as $\sin(\alpha)$; the curves are produced from a time average over
21~time frames spanning $\unit[20]{fs}$. In the collisionless case, we
have ${P}_{\perp}(x)={E}_{\perp}(x)$ and the phase shift angle
$\alpha\simeq0$ throughout the first
$\simeq\unit[25]{nm}\approx8l_{\rm s}$. Due to a finite spread in the
electron transverse velocities, there will be a continuous exchange of
electrons in the longitudinal direction not accounted for in the fluid
description above, which induces a small deviation from $\alpha=0$ and
hence $S\neq0$. However, $S$ changes sign at $x\approx\unit[15]{nm}$,
beyond which the absorbed power is negative. %
In the collisional case, ${P}_{\perp}(x)$ is consistently smaller than
${E}_{\perp}(x)$. Furthermore, the phase shift $\sin(\alpha)$ is much
larger, which is reflected in the about two orders of magnitude larger
absorbed power $S$ than in the collisionless case.

A final note on the collisional case in figure~\ref{fig:perp} (left)
is the numerical artefact that causes both ${E}_{\perp}(x)$ and
${P}_{\perp}(x)$ to level off near $x=\unit[20]{nm}$. As the Monte
Carlo collisional algorithm used in
\Smilei{}~\citep{Perez-etal_PoP2012} only conserves momentum
statistically, a ${P}_{\perp}$ noise floor is generated which drives
noise in ${E}_{\perp}$, i.e., the base level in figure~\ref{fig:perp}
(left). This effect could be alleviated by increasing the number of
macro-particles. However, the absolute majority of the collisionally
induced laser-energy absorption occurs in the region
$x=\unit[5{-}15]{nm}$ and is therefore not significantly affected by
the collisional noise floor.

\subsection{Parameter scans}
We have also performed parameter scans in order to investigate the
dependencies of the collisional heating mechanism. One such scan has
been in ionization, with either fixed ionization ($Z^*=11, 19, 24$ and
$27$) or self-consistent field and impact ionization. We have also
conducted scans in laser intensity with $a_0$ ranging from~1 to~14,
and pulse durations from $t_{\FWHM}=\unit[15]{fs}$
to~$\unit[400]{fs}$. The remaining parameters are as in
\S~\ref{sec:simulation}.

\begin{figure}
\centering
\includegraphics{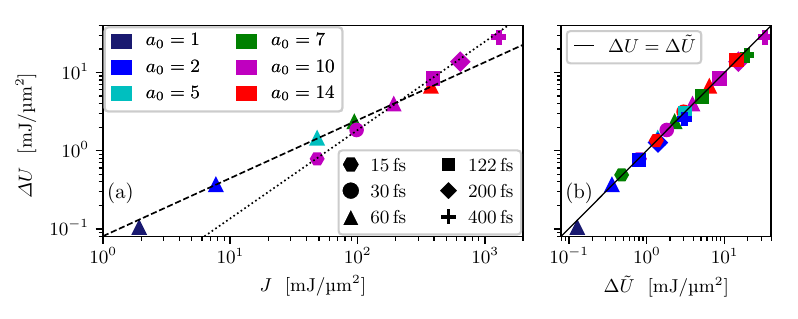}
\caption{Total simulated kinetic energy gain $\rmDelta{U}$ against (a)
  the laser energy $J$ and (b) the power law scaling
  $\rmDelta{\tilde{U}}$ for different combinations of laser parameter
  $a_0$ and duration $t_{\FWHM}$. Lines in panel~(a) indicate power
  law scalings: $\rmDelta{U}\propto J^{0.74}\propto {a_0}^{1.48}$ at
  constant $t_{\FWHM}=\unit[60]{fs}$ (dashed) and
  $\rmDelta{U}\propto J^{1.13}\propto {t_{\FWHM}}^{1.13}$ at constant
  $a_0=10$ (dotted). These two power law scalings combine to give
  $\rmDelta\tilde{U}=\unit[0.23]{mJ/\micro{m}^2}
  \times{a_0}^{1.48}\times(t_{\FWHM}/\unit[100]{fs})^{1.13}$, which is
  shown in panel~(b) to agree well with the full data set, also
  including parameter combinations not shown in~(a).  }
\label{fig:energy_scan}
\end{figure}

Let us first consider the absorbed energy from the laser.
Figure~\ref{fig:energy_scan}(\textit{a}) shows the kinetic energy gain
by the electrons and ions ($\rmDelta{U}$) following the laser
irradiation, for a scan in $a_0$ (colour coded) and a scan in pulse
duration (shape coded). The value displayed on the horizontal axis is
the laser pulse energy
$J=It_{\FWHM}[\pi/\log(4)]^{1/2}\propto{a_0}^{2}t_{\FWHM}$, where $I$
is the laser intensity. %

In the case of a constant pulse duration, $t_{\FWHM}=\unit[60]{fs}$
(triangles), the trend scales like a power law with
$\rmDelta{U}\propto J^{0.74}$ (dashed line) or
$\rmDelta{U}\propto {a_0}^{1.48}$, since $J\propto {a_0}^2$. In other
words, the absorption \emph{efficiency} scales as
$\rmDelta{U}/J\propto J^{-0.26}\propto {a_0}^{-0.52}$. %
The $a_0=1$ point seems to deviate from the above scaling, suggesting
that it is mainly valid at relativistic intensities.  Thus, $a_0=1$
was excluded from the fit. %
This scaling is similar to the ${I}^{-1/4}$ scaling of the normal skin
effect, as described by \citet{Rozmus-Tikhonchuk_PRA1990}.  Comparing
the collisional mean free path, $\lambda_{\rm mfp}\sim\unit[20]{nm}$,
to the skin depth, $l_{\rm s}\approx\unit[6]{nm}$\footnotemark{}, it
is not completely clear that the normal skin effect
($\lambda_{\rm mfp}\lesssim{}l_{\rm s}$) can be ruled out. However,
our simulation results do not support some other scaling laws derived
by \citet{Rozmus-Tikhonchuk_PRA1990} for normal skin effect, possibly
due to the non-relativistic and simplified nature (constant intensity
and semi-infinite plasma) of their analytic treatment.
\footnotetext{The skin depth adjusted for collisions has been inferred
  from figure~\ref{fig:perp} (left).}

The other scaling trend displayed in
figure~\ref{fig:energy_scan}(\textit{a}) is at constant $a_0=10$
(magenta). Here, the power law fit (dotted line) gives
$\rmDelta{U}\propto J^{1.13}\propto {t_{\FWHM}}^{1.13}$. In this case
the absorption efficiency still has a weak positive scaling of
$\rmDelta{U}/J\propto {t_{\FWHM}}^{0.13}$. %
From the pulse duration scaling, we note that the $\unit[200]{fs}$ and
$\unit[400]{fs}$ pulses begin to fall off below the scaling followed
by the other data points, and they were thus also excluded from the
fit. The decreasing trend with $t_{\FWHM}$ in the long-pulse limit may
be a consequence of the recirculation of hot electrons, which could
lower the effective plasma collisionality in the irradiated region.

Combining the two above scalings yields the approximate scaling
\begin{equation}
\rmDelta\tilde{U}=\unit[0.23]{mJ/\micro{m}^2}\times
{a_0}^{1.48} \qty(\frac{t_{\FWHM}}{\unit[100]{fs}})^{1.13}
\end{equation}
in the $(a_0,t_{\FWHM})$ plane.
Figure~\ref{fig:energy_scan}(\textit{b}) shows this scaling to agree
well with the full set of data, including data points where both $a_0$
and $t_{\FWHM}$ are varied. %
The observed scaling does break down at the low-$a_0$ or long-duration
limits. However, the range of validity stretches over two orders of
magnitude in pulse energy and a similar range in absorbed energy, and
the laser parameters captured by this power law are experimentally
feasible and relevant to isochoric heating experiments. %
Collisional effects in general decrease at higher particle energies
and the absorption happens through collisional scattering of the
laser-driven electrons in the skin layer. Therefore, if the laser
field ($a_0$) is increased, so that the electrons in the skin layer
reach higher energies, then the efficiency of the collisional
absorption should decrease.

\begin{figure}
\centering
\includegraphics{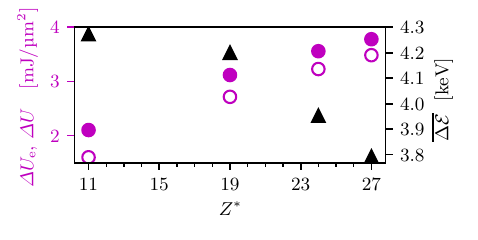}
\caption{Collisional simulation with various (fixed) ionization degree
  $Z^{*}$. Left axis (magenta circles): gained kinetic energy by
  electrons \emph{and} ions $\rmDelta{U}$ (filled circles) as well as
  only by electrons $\rmDelta{U}_{\ee}$ (open circles). Right axis
  (black triangles): average kinetic energy gained by one electron
  $\overline{\rmDelta\EE}$. }
\label{fig:Zstar_scan}
\end{figure}

We also report on a scan in (fixed) ionization degree $Z^{*}$.
Although this parameter cannot be controlled independently in
experiments, this scan aims to provide insight into the target
collisionality, which scales as $(Z^{*})^2$~--~ignoring other effects,
e.g.\ individual particle energy. However, by varying $Z^{*}$ while
keeping the ion density $n_{\Cu,0}$ fixed, we inevitably also change
the electron density $n_{\ee,0}=Z^{*}n_{\Cu,0}$, which may introduce
other density-related effects. Nevertheless, the electron density
stays highly overcritical~--~the lowest electron density in this scan
is $n_{\ee,0}=532.4\nc$ for $Z^{*}=11$. %
Figure~\ref{fig:Zstar_scan} displays the energy absorbed by both
electrons and ions $\rmDelta{U}$ (filled circles, left axis) and only
by electrons $\rmDelta{U}_{\ee}$ (open circles, left axis) for the
different ionization degrees. The absorbed energy increases with
$Z^{*}$, while a smaller fraction of the absorbed energy goes into the
ions at higher $Z^{*}$.

Due to the accompanying changes in electron density, the average
absorbed energy \emph{per electron} $\overline{\rmDelta\EE}$, also
shown in figure~\ref{fig:Zstar_scan} (black triangles, right axis),
happens to decrease by approximately $15\,\%$ from $Z^{*}=11$ to $27$.
The decrease in $\overline{\rmDelta\EE}$ with $Z^{*}$ may seem
surprising if the dominant heating mechanism is collisional. However,
this might be due to other density effects, such as the increased skin
depth at lower electron density which allows a deeper laser
penetration and thus a stronger laser-to-electron coupling
efficiency. When we examine the electron energy spectra (not shown) in
this scan, the Maxwellian-fitted bulk electron temperatures are all
$T_{\ee}=\unit[3.5{\pm}0.1]{keV}$.

\begin{figure}
\centering
\includegraphics{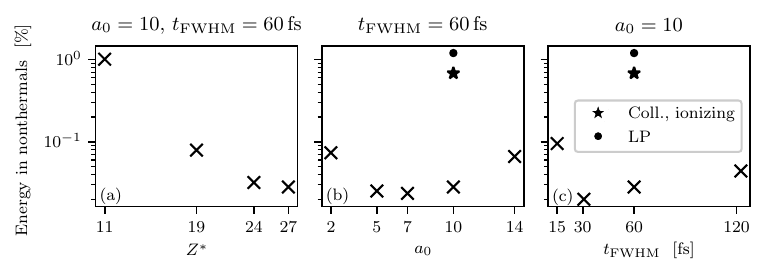}
\caption{Fraction of the electron energy in non-thermal electrons, for
  collisional simulations, $\unit[200]{fs}$ after the end of the laser
  pulse, for scans in $Z^{*}$~(a), $a_0$~(b) and $t_{\rm FWHM}$~(c)
  with CP, marked by crosses. The value marked with a dot is from LP,
  and self-consistent ionization is shown as a star. %
}
\label{fig:non-thermal}
\end{figure}

Besides just the pure amount of energy absorbed from the laser, we are
also interested in how well thermalized the plasma is. As a measure of
that, figure~\ref{fig:non-thermal} shows the fraction of electron
kinetic energy in the non-thermal electrons. %
This is calculated by the fraction of the energy in the high-energy
tail to the total electron kinetic energy, %
$\int_{2T_{\ee}}^{\infty} %
[f_{\EE}(\EE)-f_{\EE}^{\rm MJ}(\EE)]\EE\,\dd\EE$, %
where $f_{\EE}^{\rm MJ}$ is a Maxwell--J\"{u}ttner distribution fitted
to the bulk of the electron energy spectrum $f_{\EE}(\EE)$ (as shown
in figure~\ref{fig:energy-spectrum}) and $T_{\ee}$ is the temperature
inferred from the fit.
Figure~\ref{fig:non-thermal} shows scans in fixed ionization degree
$Z^{*}$~(a), laser amplitude $a_0$~(b) and pulse duration
$t_{\rm FWHM}$~(c). The displayed values are taken $\unit[200]{fs}$
after the end of the laser pulse. Due to a varying heat transport
speed, the fraction of non-thermal electron energy is only taken in
the region in which the bulk electron temperature is no longer
increasing. This should still give a representative estimate of the
non-thermal fraction, since the fast electrons have already
recirculated by the chosen time, see the $t=\unit[250]{fs}$ panels of
figure~\ref{fig:electron_dist}. The precise values in
figure~\ref{fig:non-thermal} are sensitive to the choice of time and
region to include, thus these results are only
qualitative. Nevertheless, the general trends shown here are still
representative of the observed situation~--~importantly, the relation
between LP and CP is robust.

Figure~\ref{fig:non-thermal}(\textit{a}) shows that there is a trend
toward lower non-thermal fraction at higher ionization levels, which
is consistent with the faster thermalization expected at high
$Z^*$. This trend also suggests that the higher absorbed energy per
electron at lower $Z^{*}$ (figure~\ref{fig:Zstar_scan}) is linked to a
relative increase in the non-thermal population.

Regarding the scans in the laser amplitude and duration in
figure~\ref{fig:non-thermal}(\textit{b}) and
\ref{fig:non-thermal}(\textit{c}), respectively, no clear trend
appears to emerge among the CP laser pulses (marked by $\times$).
Then there are the self-consistent ionization (star) and LP (downward
triangle) simulations: both have approximately one order of magnitude
higher fraction of energy in non-thermal electrons than the equivalent
(fixed-ionization, CP) counterpart. The higher fraction of non-thermal
energy with LP stems from the \jxB{} and vacuum heating mechanism. The
higher non-thermal energy fraction with a self-consistent ionization
process is discussed in conjunction with its phase-space distribution
in figure~\ref{fig:electron_dist}.

Even a very small fraction of non-thermals may affect the
interpretation of X-ray
diagnostics~\citep{Rosmej_JPB1997,Chen_PoP_2009,Renner-Rosmej_MRE2019},
meaning that LP can be more intrusive than CP in WDM/HDM studies.  We
have also conducted simulations with a larger pre-expanded plasma
(exponentially decaying density profile with a scale length of
$\unit[80]{nm}$). While not presented here, those simulations show
that LP can result in up to ${\sim}10\%$ of the electron kinetic
energy in non-thermal electrons, which would of course be even more
intrusive and significantly affect the X-ray diagnostics. With CP, the
pre-plasma weakens the energy absorption by about a factor of two, but
the fraction of energy in fast electrons stays $\lesssim1\%$.

\subsection{Two-dimensional simulation results}

\begin{figure}
\centering
\includegraphics{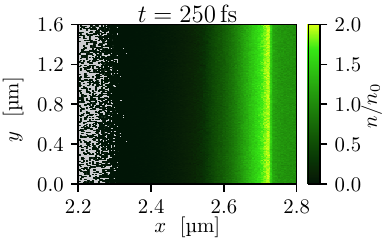}
\put(-57,30){(a)}
~
\includegraphics{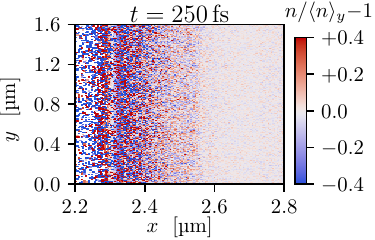}
\put(-56,30){(b)}
\\
\includegraphics{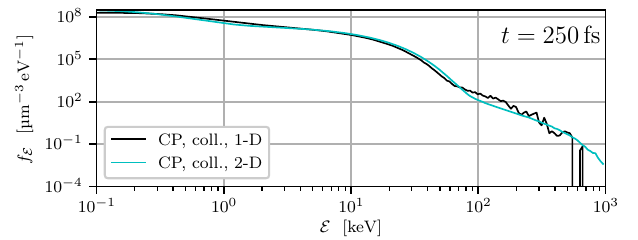}
\put(-50,30){(c)}
\caption{Map of the electron density (a) in units of background
  density $n_0$ and the relative transverse variation of the electron
  density (b) near the front target surface at $\unit[250]{fs}$. Panel
  (c) shows a comparison of the electron energy spectra from the 1-D
  and 2-D simulations at the same time. The spectra are taken from the
  full simulation box.  The transverse band of higher density at
  $x\approx\unit[2.72]{\micro{m}}$ in the panel (a) corresponds to a
  shock front launched by the laser impact. }
\label{fig:2D}
\end{figure}

Up to this point, all the results presented have been produced in 1-D
simulations. However, to investigate the applicability of these
results in higher dimensions, where transverse plasma modulations can
arise at the target boundary and affect the bulk heating and
hot-electron generation~\citep{Kemp-Divol_PoP2016}, we have performed
a 2-D simulation at our baseline laser parameters (see
\S~\ref{sec:simulation} for details).
A density map of the electron density near the illuminated target
surface is shown in figure~\ref{fig:2D}(\textit{a}). The density is
shown at $t=\unit[250]{fs}$. There is a transverse band of higher
density at $x\approx\unit[2.72]{\micro{m}}$ that represents a shock
front propagating into the plasma. Notably, this shock front remains
straight, with no evidence of substantial density modulations.

To more clearly demonstrate the absence of transverse instability
effects, the relative transverse variation of the electron density
$n/\ev{n}_{y}-1$ is displayed in figure~\ref{fig:2D}(\textit{b}),
where $\ev{\cdot}_{y}$ denotes a transverse average over the full
width of the simulation box in $y$. Any transverse density structure
should therefore be clearly visible in this representation.  The
substantial deviations from the average density observed in the
low-density ($n_e \lesssim 0.2n_0$) region correspond to statistical
noise due to a low number of computational particles in said
region. Importantly, the deviations seen in
figure~\ref{fig:2D}(\textit{b}) have no structure to them, and the
same applies for the shock, suggesting that transverse effects are
inoperative in the present highly collisional case (at least within
the simulated time window).

Lastly, to confirm that the collisional heating behaves similarly in
the 2-D and 1-D simulations, figure~\ref{fig:2D}(\textit{c}) shows the
electron kinetic energy spectra of the corresponding 1-D and 2-D
simulations at $t=\unit[250]{fs}$.\footnote{%
  Due to the 2-D simulation having a smaller longitudinal box size,
  and thus the target front being located at $x=\unit[2.5]{\micro{m}}$
  instead of $x=\unit[7.5]{\micro{m}}$, the times of comparison for
  the \mbox{1-D} simulation are shifted by $\unit[15]{fs}$ later
  compared to the 2-D simulation, due to increased travel time for the
  laser pulse. The simulation time at which the 1-D spectrum is
  plotted is thus $t=\unit[265]{fs}$.} %
The spectra of the \mbox{2-D} and \mbox{1-D} simulations are
essentially the same. Although not shown here, the two spectra agree
similarly well also at earlier times. Since the following
thermalization process is almost entirely collisional, and thus
independent of dimensionality, it is safe to conclude that the energy
absorption is not affected by going from one to two dimensions, under
the interaction conditions considered. %

\section{Conclusions}

We have performed collisional and collisionless 1-D and 2-D PIC
simulations and shown that a collisional, inverse bremsstrahlung,
absorption can be used for strong plasma heating in a solid-density,
high-$Z^{*}$ material, such as copper, with ultrahigh intensity,
short-pulse lasers. Using CP, the electron population quickly
thermalizes to well-formed Maxwellian distributions suitable for
experimental verification of HED physics models. %
The collisional simulations show that the target electrons are quickly
heated to $T_{\ee}\sim\unit[3.5]{keV}$ bulk temperature on a
${\sim}\unit[300]{fs}$ time scale. The target energy density reaches
${\sim}\unit[10]{Gbar}$, %
which is within the realm of ultrahigh energy density. The use of CP
provides faster collisional thermalization of the electron population
compared to LP, something which is valuable for experimental tests of
HED atomic physics models. %
A test of the collisional absorption using a 2-D simulation,
demonstrates the transferability of the \mbox{1-D} results to higher
dimensions. In contrast to previous work conducted with lower-$Z^*$
targets~\citep{Kemp-Divol_PoP2016}, the high collisionality is not
favourable for driving transverse plasma modulations, resulting in the
same absorption levels in two dimensions as in one dimension, at least
for CP. %

We have carried out scans over laser parameters and ionization. The
scans over laser settings show that the mechanism is robust to changes
in the laser, over two orders of magnitude in laser energy, with lower
intensity and longer pulses at the same laser pulse energy yielding
better energy absorption. Also, the higher collisionality incurred
from a higher ionization level improves energy absorption and electron
thermalization. %
A more realistic simulation run with self-consistent ionization,
including both impact and field ionization, reached
$T_{\ee}\sim\unit[2.5]{keV}$, confirming that collisional heating is
still operational in a self-consistently ionized plasma, although its
thermalization is then less complete than at fixed ionization, due to
high-energy electrons generated through ionization events in
strong-field regions.

%_________________________________________________
%-------------------------------------------------
%_________________________________________________
%% ACKNOWLEDGEMENTS
\acknowledgements %
The authors are grateful for fruitful discussions with L. Hesslow and
T. F\"{u}l\"{o}p, as well as to M. Grech and F. P\'{e}rez for support
with \Smilei{}.
This project has received funding from the European Research Council
(ERC) under the European Union's Horizon 2020 research and innovation
programme under grant agreement no. 647121, the Swedish Research
Council (grant no. 2016-05012), and the Knut och Alice Wallenberg
Foundation.
The simulations were performed on resources provided by the Swedish
National Infrastructure for Computing (SNIC) at Chalmers Centre for
Computational Science and Engineering (C$^3$SE) and High Performance
Computing Center North (HPC$^2$N).

%%%%%%%%%%%%%%%%%%%%%%%%%%%%%%%%%%%%%%%%%%%%%%%%%%%%%%%%%%%%%%%%%%%%%%
%\appendix
%%%%%%%%%%%%%%%%%%%%%%%%%%%%%%%%%%%%%%%%%%%%%%%%%%%%%%%%%%%%%%%%%%%%%%

%%%%%%%%%%%%%%%%%%%%%%%%%% The bibliography %%%%%%%%%%%%%%%%%%%%%%%%%%
\bibliographystyle{jpp}
\bibliography{references}%requires a file named 'references.bib'
%%%%%%%%%%%%%%%%%%%%%%%%%%%%%%%%%%%%%%%%%%%%%%%%%%%%%%%%%%%%%%%%%%%%%%

%%%%%%%%%%%%%%%%%%%%%%%%%%%%%%%%%%%%%%%%%%%%%%%%%%%%%%%%%%%%%%%%%%%%%%
\end{document}